\documentclass[twocolumn,tighten,trackchanges]{aastex631}
\usepackage{amsmath}
\hypersetup{colorlinks=true,linkcolor=blue,citecolor=blue,filecolor=cyan,urlcolor=blue}
\shorttitle{Subsurface Flows and Magnetic Flux Transport}
\shortauthors{Sen et al. }
\graphicspath{{./}{figures/}}

\begin{document}

\title{Active-region modulation of subsurface meridional flows and magnetic flux transport on the Sun}

\author[0000-0003-2694-3288]{Anisha Sen} 
\email{anisha.sen@iiap.res.in}
\affiliation{Indian Institute of Astrophysics, II Block, Koramangala, Bengaluru 560 034, India}
\affiliation{Pondicherry University, R.V.Nagar, Kalapet, Puducherry 605014, India}

\author[0000-0003-0003-4561]{S.P. Rajaguru}
\email{rajaguru@iiap.res.in}
\affiliation{Indian Institute of Astrophysics, II Block, Koramangala, Bengaluru 560 034, India}
\affiliation{Pondicherry University, R.V.Nagar, Kalapet, Puducherry 605014, India}


\author[0000-0002-2632-130X]{Ruizhu Chen}
\affiliation{W. W. Hansen Experimental Physics Laboratory, Stanford University, Stanford, CA 94305-4085, USA}

\author[0000-0002-6308-872X]{Junwei Zhao}
\affiliation{W. W. Hansen Experimental Physics Laboratory, Stanford University, Stanford, CA 94305-4085, USA}

\author[0000-0003-1860-3697]{Shukur Kholikov}
\affiliation{National Solar Observatory, Boulder CO, USA}









\begin{abstract}

Using time-distance helioseismology applied to 14-years of SDO/HMI observations spanning solar cycle 24 and rising 
phase of cycle 25, we present evidence that meridional flows in the lower half of the near-surface shear layer (NSSL),
modulated by active-region magnetic fields, play a central role in the episodic global transport of magnetic 
flux. In particular, polar field buildup is tightly linked to plasma outflows diverging from active latitudes 
within the deeper NSSL. The magnitude and timing of hemispheric polar field evolution are regulated by depth-dependent 
meridional flow, including its cross-equatorial component, responding to active-region flux asymmetries.
During cycle 24 maximum, stronger southern outflows accelerated flux transport, causing the southern polar field 
to peak nearly four years before the northern. Global magnetic flux transport patterns in the previous 
three solar cycles (21, 22, and 23) show broad consistency with the deeper meridional flow modulation inferred
in cycles 24 and 25. These results identify activity-dependent flow variations in deeper layers of the NSSL as a 
dynamically significant component of the Babcock-Leighton process that governs the generation and hemispheric 
asymmetry of global dipole field.

\end{abstract}


\section{Introduction} 
\label{sec:intro}

The physical processes that control the transport and redistribution of solar photospheric magnetic field,
collectively described as surface flux transport (SFT)\citep{2023SSRv..219...31Y}, form the crucial closing loop 
producing large-scale poloidal field for the cyclic operation of solar dynamo in a class of models known as
the Babcock-Leighton Flux Transport (BLFT) dynamos \citep{2014ARA&A..52..251C}. Over the years since its inception \citep{
1961ApJ...133..572B,1964ApJ...140.1547L,1969ApJ...156....1L}, the BLFT models have been 
developed incorporating key observed phenomena behind the near-surface evolution of magnetic fields.
A major development has been the inclusion of large-scale meridional circulation in the BLFT models 
providing ways to explain several observed features of the solar cycle \citep{1991ApJ...375..761W,1995A&A...303L..29C}. 
Currently, this area of research has seen intense efforts to develop observational data-driven  
modelling of evolution of magnetic fields due to both mean flows and turbulent convection, and their variability
linked to the emergence and formation of magnetic structures themselves and thus acting as self-regulating
feedback loop \citep{2010ApJ...717..597J,2010ApJ...720.1030C,2011Natur.471...80N,2012A&A...548A..57C,
2014ApJ...785L...8M,2015ApJ...808L..28J,2023SSRv..219...39H,2024SoPh..299...42T}.

Following early observational attempts to measure polar magnetic fields and their cyclic variation \citep{1955ApJ...121..349B,
1959ApJ...130..364B,1978SoPh...58..225S}, \citet{2005GeoRL..32.1104S} and \citet{2013ApJ...763...23S} demonstrated 
that the strength of the polar fields near solar minimum is a reliable precursor for predicting the amplitude of the 
subsequent solar cycle, highlighting the central role of polar-field buildup in the solar dynamo process. 
\cite{2008ApJ...688.1374T} used high-resolution Hinode/SOT spectropolarimetry to show that the Sun’s polar 
regions contain numerous 1 kG unipolar vertical magnetic flux tubes of the same polarity. Analysis of long-term
synoptic data by \citet{2010SoPh..267..267J} revealed a pronounced drop in polar field strength from the late 
decline of cycle 22 into cycle 23, tightly correlated with changes in the meridional flow speed (calculated by 
\citet{2010Sci...327.1350H}). Modeling studies by \citet{2010ApJ...717..597J} revealed that activity-related 
converging inflows toward active latitudes can reduce the peak polar-field strength by approximately 18\%, while 
\citet{2014ApJ...780....5U} demonstrated that meridional flow variations during cycle 23 led to a 20\% reduction 
in polar-field amplitude in their data-assimilative SFT model. \citet{2011A&A...528A..83J} used 
a SFT model with reconstructed sunspot emergence records to successfully reproduce polar and 
open flux back to 1700, confirming the strong correlation between polar field at activity minimum and the 
strength of the subsequent sunspot cycle. Further theoretical investigations by \citet{2011Natur.471...80N} and 
\citet{2012A&A...548A..57C} showed that localized meridional-flow perturbations impede poleward magnetic flux 
transport, weakening the polar fields and strength of subsequent solar cycle. \citet{2020ApJ...904...62W} analyzed 
how persistent activity complexes during cycle 24 generated a substantial poleward magnetic flux surge, 
thereby significantly influencing the evolution of the Sun’s polar magnetic fields. More recent high-resolution 
observations by \citet{2024RAA....24g5015Y} confirm that latitude-dependent variations in poleward flow can create 
hemispheric asymmetries in the timing of polar-field reversal. 

Based on solar-cycle-long observations of meridional and zonal flows in the Sun’s near-surface shear layer 
(NSSL: \citet{1996Sci...272.1300T}), \citet{2025ApJ...984L...1S} showed that near-surface inflows toward active 
latitudes are part of localized circulation cells around active regions, with outflows at depths below $\approx$ 0.97$R_{\sun}$, 
coinciding with changes in radial gradient of rotation. A follow-up study by \citet{2026ApJ...997...57S}
demonstrated that the equatorward part of such outflows drive cross-equatorial flows from the more active hemisphere,
facilitating flux cancellation near the equator.

In this Letter, we report results from a thorough investigation of the relationships between the surface flux transport from the active
latitudes and the outflows originating beneath them in lower half of the NSSL in cycle 24 and rising phase of cycle 25. 
We also examine the synoptic magnetic fields of previous three cycles (21 to 23) for deeper flow-driven signatures in
surface flux transport, including those in cross-equatorial flux plumes driven by hemispheric asymmetry in active region 
magnetic fields. The Letter is structured as follows: Section \ref{sec: data} outlines 
the data utilized and a description of the analysis technique. Section \ref{sec: result} presents our findings, 
and Section \ref{sec: conclusion} discusses and concludes outlining the broader implications of our results.


\section{Data and Analysis Procedure} \label{sec: data}

We employ identically processed helioseismic observations obtained from the space-based Helioseismic and Magnetic Imager
(HMI: \citealt{scherrer2012helioseismic}) onboard NASA’s Solar Dynamics Observatory (SDO) and from the ground-based
Global Oscillation Network Group (GONG). Time-distance helioseismology \citep{duvall1993time} is performed with these datasets
to infer the meridional flow \citep{rajaguru2015meridional} structure and dynamics within the near-surface shear layer (NSSL). 
The data sources and analysis approach used in this study are the same as those described by \citet{2025ApJ...984L...1S}, 
to whom we refer the readers for details and for a discussion on agreement between HMI and GONG measurements.
For the analysis and presentation here, we use only the HMI data. The analysis spans a 14-year period from May 2010 to 
April 2024, one year longer into Cycle 25 than used in \citet{2025ApJ...984L...1S}. The measurements were processed at a 
reduced spatial resolution of 0.36° per pixel. To mitigate magnetic 
contamination in the flow estimates \citep{2015ApJ...805..165L,2017ApJ...849..144C}, regions with line-of-sight magnetic 
field strengths exceeding 40 G were masked out in the Dopplergrams prior to analysis. Meridional flows were derived by 
inverting for the stream function, ensuring compliance with the continuity equation and, consequently, the conservation of mass 
\citep{rajaguru2015meridional}. 

For studying the connection between flows and the temporal and 
latitudinal evolution of sunspot distributions, we use data from the Solar Region Summary provided by the 
National Oceanic and Atmospheric Administration (NOAA). 
We have also collected synoptic magnetogram data, extending back to three solar cycles, from multiple sources: data 
from the National Solar Observatory, Kitt Peak, USA (NSO/Kitt Peak; \url{https://nispdata.nso.edu/ftp/kpvt/synoptic/mag/}) 
cover Carrington Rotations (CR) 1625-2007; this was followed by data from the Vector Spectromagnetograph (VSM) 
of the NSO/Synoptic Optical Long-term Investigations of the Sun (SOLIS; \url{https://solis.nso.edu/0/vsm/crmaps/}) for 
CR 2008-2096, and subsequently data from the SDO/HMI (\url{http://jsoc.stanford.edu/}) for CR 
2097-2301. The complete dataset spans the period from 19 February 1975 to 9 September 2025. Data for eight 
Carrington Rotations (CRs 1640, 1641, 1642, 1643, 1644, 1854, 2015, and 2041) were unavailable and were filled by 
interpolation for continuity in the analysis.

A known limitation of above described synoptic magnetic maps arises near the solar poles due to projection effects. 
Hence, for more accurate polar field measurements, we used data from the Wilcox Solar Observatory (WSO; 
\url{http://wso.stanford.edu/Polar.html} ), where the polemost aperture measures the line-of-sight field from 
approximately 55° latitude to the poles. Monthly averaged WSO data from 31 May 1976 to 11 September 2025 were 
employed, with four missing months (September 2022, October 2022, February 2024, and May 2025) filled by 
interpolation. For a consistent comparison between the synoptic and WSO polar field datasets, the common time 
interval from 31 May 1976 to 11 September 2025 was used.

\graphicspath{ {Image/} }
\begin{figure*}[!htbp]
\gridline{\fig{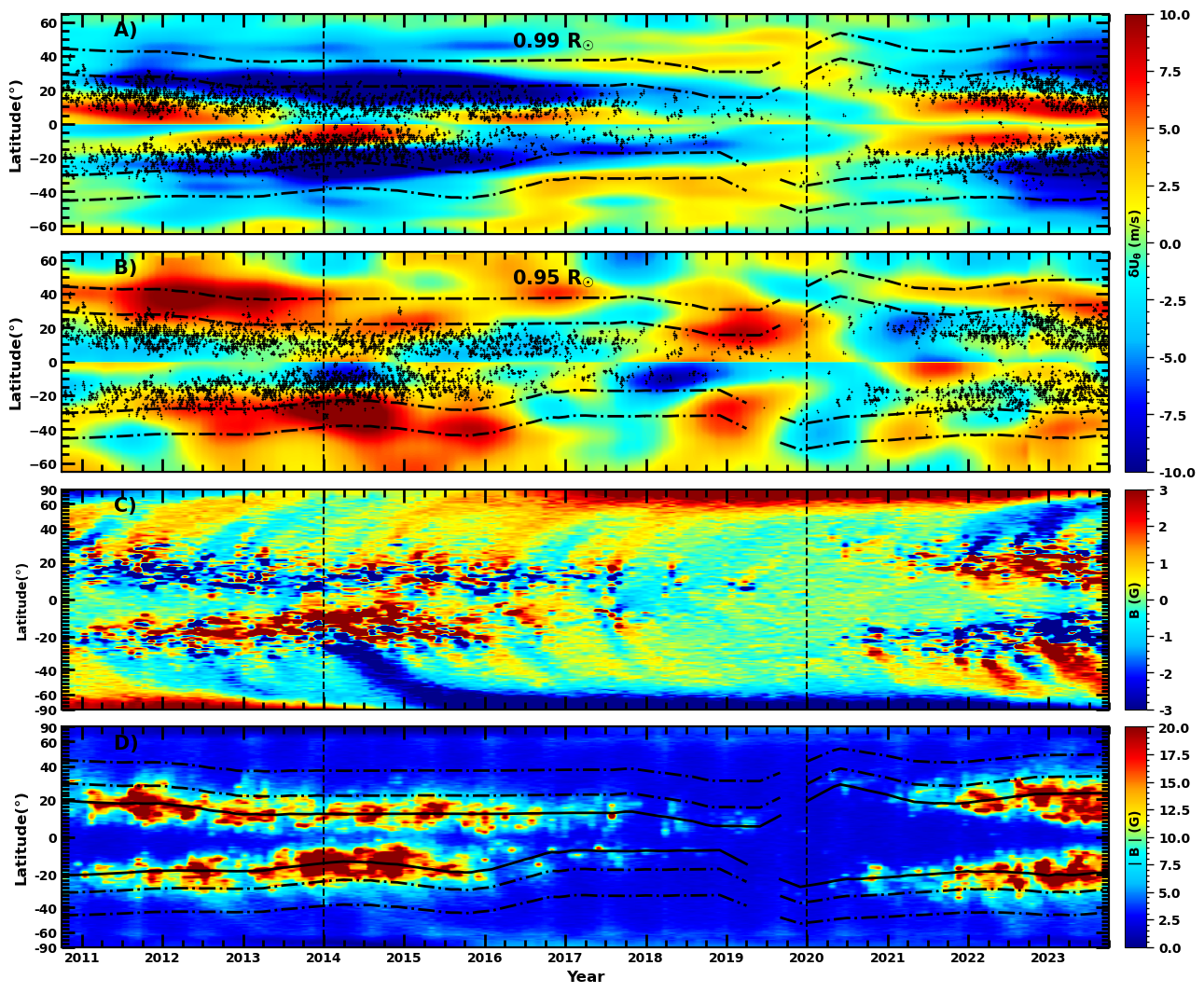}{1\textwidth}{}
          }
\caption{
Time-latitude profile of variations in the residual meridional flow at two depths, 0.99 $R_{\sun}$ and 0.95 $R_{\sun}$,
respectively are in panels A) and B). Positive values correspond to poleward flow in both hemispheres. Corresponding 
time-latitude variations in the longitudinally averaged signed (\textit{i.e.}, 
the magnetic butterfly diagram) and unsigned magnetic field are in the lower panels C) and D), respectively. 
The solid line overplotted in panel D marks mean latitude, $\theta_{m}$, of peak magnetic flux in each hemisphere, and the 
dot-dashed lines in panels A, B and D enclose the region 10° to 25° away from $\theta_{m}$. The two vertical 
dotted lines mark epochs of Solar Cycle 24 maximum (2014) and minimum (2020). Sunspot locations are overplotted as black dots.}

\label{fig:1}
\end{figure*}

\section{Results} \label{sec: result}
\subsection{Outflows Beneath Active Regions and Surface Flux Transport}
\label{subsec: results_1}



\begin{figure*}
\centering
\begin{minipage}{0.50\textwidth}
    \centering
    \includegraphics[width=\linewidth,height=10cm]{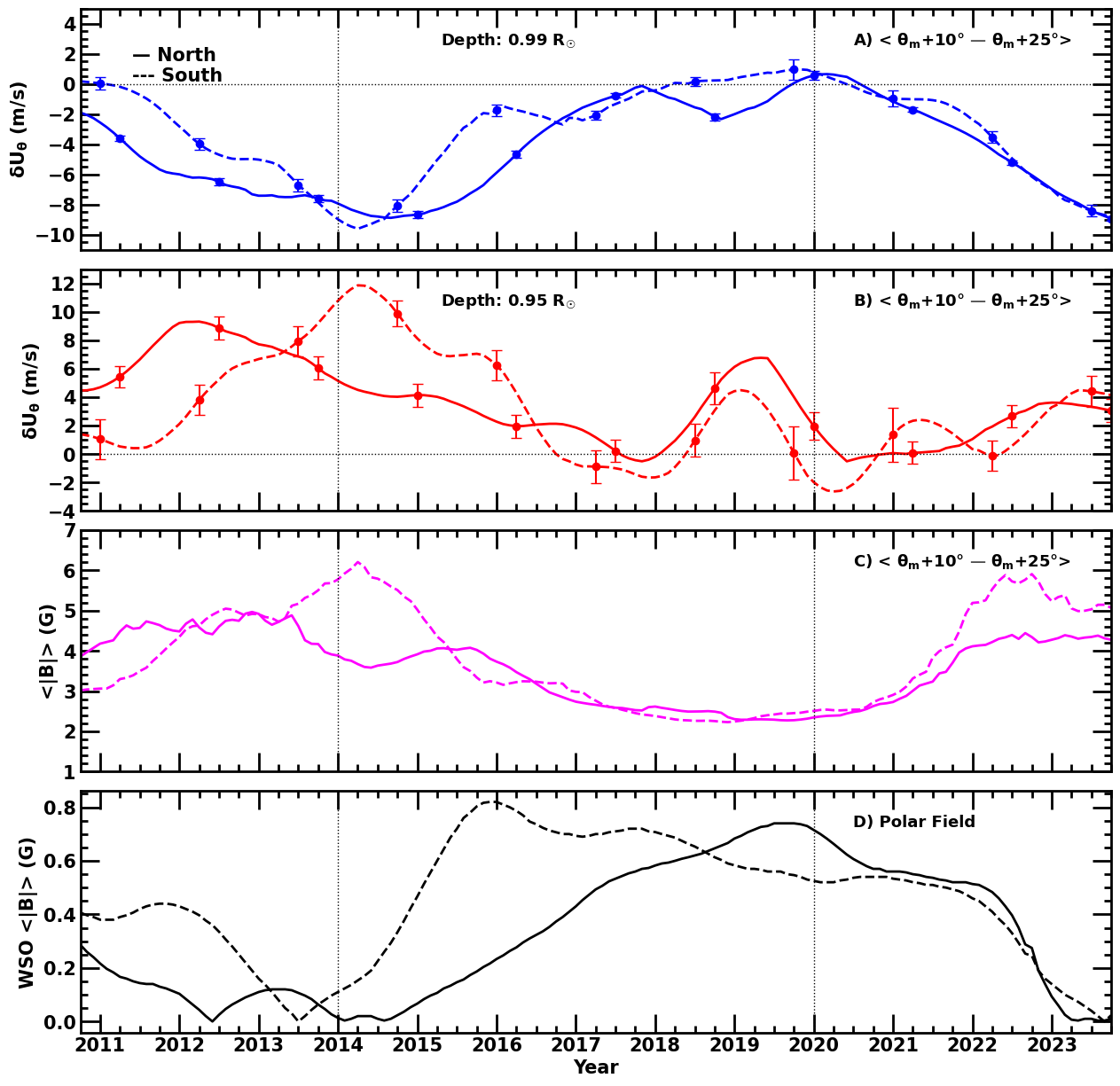}
\end{minipage}
\hfill
\begin{minipage}{0.48\textwidth}
    \centering
    \includegraphics[width=\linewidth,height=5.0cm]{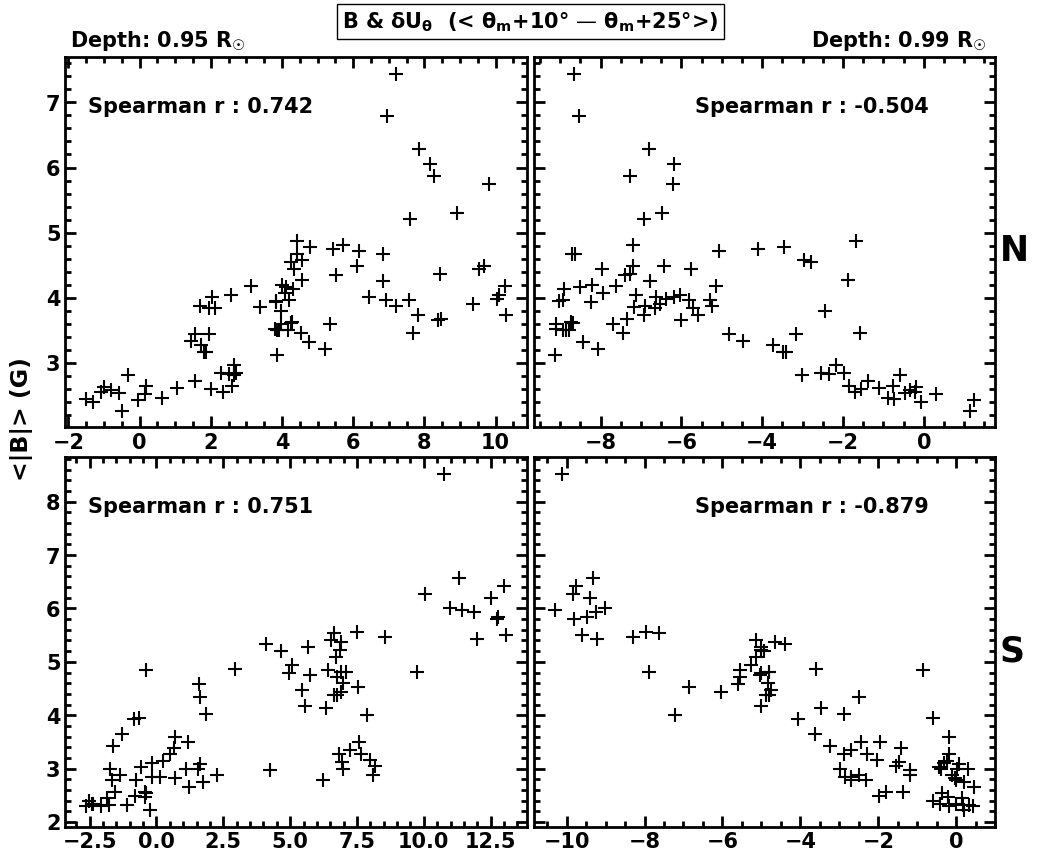}
    \vspace{0.5cm}
    \includegraphics[width=\linewidth,height=5.0cm]{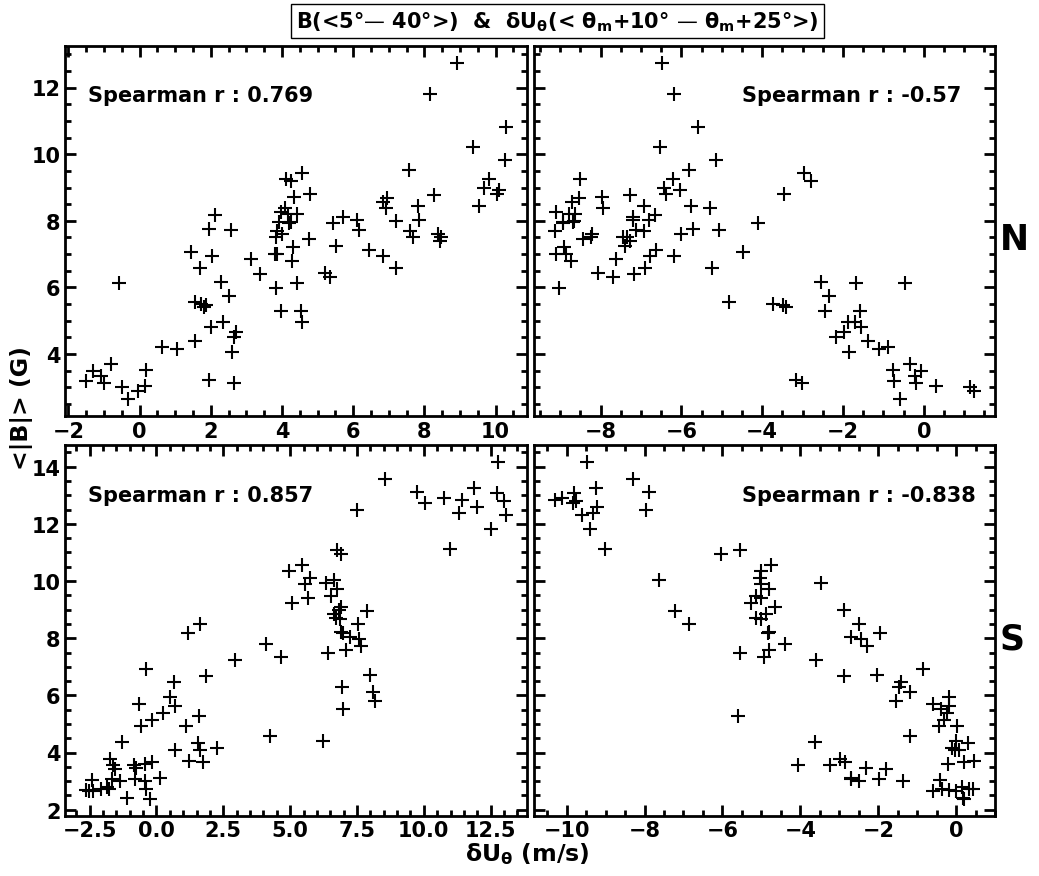}
\end{minipage}

\caption{\textbf{Left:} Panels A and B show the residual meridional flows averaged over 10°-25° away from 
$\theta_{m}$ (marked by dot-dashed lines in Figure \ref{fig:1}) in the northern (solid) and southern (dashed) hemispheres 
at depths 0.99$R_{\odot}$ and 0.95$R_{\odot}$, respectively. Panel C shows the similarly averaged unsigned LOS HMI 
magnetic field ($|B|$), while panel D shows the unsigned polar field from WSO measurements. \textbf{Right:} Scatter 
plots illustrating the correlation between variation of unsigned magnetic field and the flow residuals covering the period
of Cycle 24 (Oct 2010 - June 2018), along with the estimated Spearman rank correlation coefficients; the top four panels 
correspond to both quantities averaged over the same latitudinal range (10°-25° away from $\theta_{m}$), at depths of 0.95$R_{\odot}$ 
(left column) and 0.99$R_{\odot}$ (right column), for the northern (top row) and southern (bottom row) hemispheres, while
the bottom four panels show the correlation between same flow signals against the unsigned magnetic field averaged over
the whole active latitude range of 5°-40°.}
\label{fig:2}
\end{figure*}

In this Section, we investigate how the active-region driven local circulations within the NSSL modulate
the meridional flows and, consequently, the transport and accumulation of magnetic flux toward the poles. 
In our previous study \citep{2025ApJ...984L...1S}, we determined the global scale residuals in meridional
flow by subtracting its average over a solar cycle (see, for example \citet{antia2022changes}). Here however,
to highlight better the variations due to active region magnetic fields, we subtract baseline meridional flow
of the Sun determined as an average over the solar minimum period (about 2 years covering May 2019 - April 2021), 
when there were no active regions \citep{mahajan2023removal,2026ApJ...997...57S}. 
The top two panels of Figure \ref{fig:1} present the time - latitude profile of cycle-minimum-subtracted variations in meridional flow, 
$\delta U_{\theta}$, at depths of 0.99$R_{\sun}$ and 0.95$R_{\sun}$, covering the period from October 2010 to October 2023.
We use the sign convention of positive values for poleward flows in both hemispheres.
Measurement errors in flows are calculated by repeating the seismic inversions 1000 times with travel times randomly 
perturbed with estimated errors in observed values \citep{2025ApJ...984L...1S}. While the near-surface flows are measured 
at 5$\sigma$ level or above, those in the deeper layers (0.95$R_{\sun}$) can decrease to 3$\sigma$ levels: the outflow signals 
on the poleward side of active latitudes (between 20° and 40°) are in the range of 3 to 12 m/s, with estimated errors 
ranging from 0.3 to 1.1 m/s (see Figure \ref{fig:2}). 
The lower two panels of Figure \ref{fig:1} present the longitudinally averaged photospheric radial 
magnetic field over time and latitude -- the magnetic butterfly diagram -- for the same period as for the flows, and together 
they provide a comprehensive view of the evolution of large-scale flow and magnetic patterns across Solar Cycle 24 and into the early 
phase of Solar Cycle 25: panel C shows the signed radial magnetic field, which preserves polarity information and thus allows  
the tracking of poleward flux transport, polarity reversals, and other large-scale magnetic field evolution processes; panel D
displays the absolute magnitude of the radial magnetic field, which helps emphasizing the total flux transported as the flow itself
does not differentiate the polarity of the fields. Sunspot locations are overplotted as black dots in the panels for the flows.

\begin{figure*}[!htbp]
\gridline{\fig{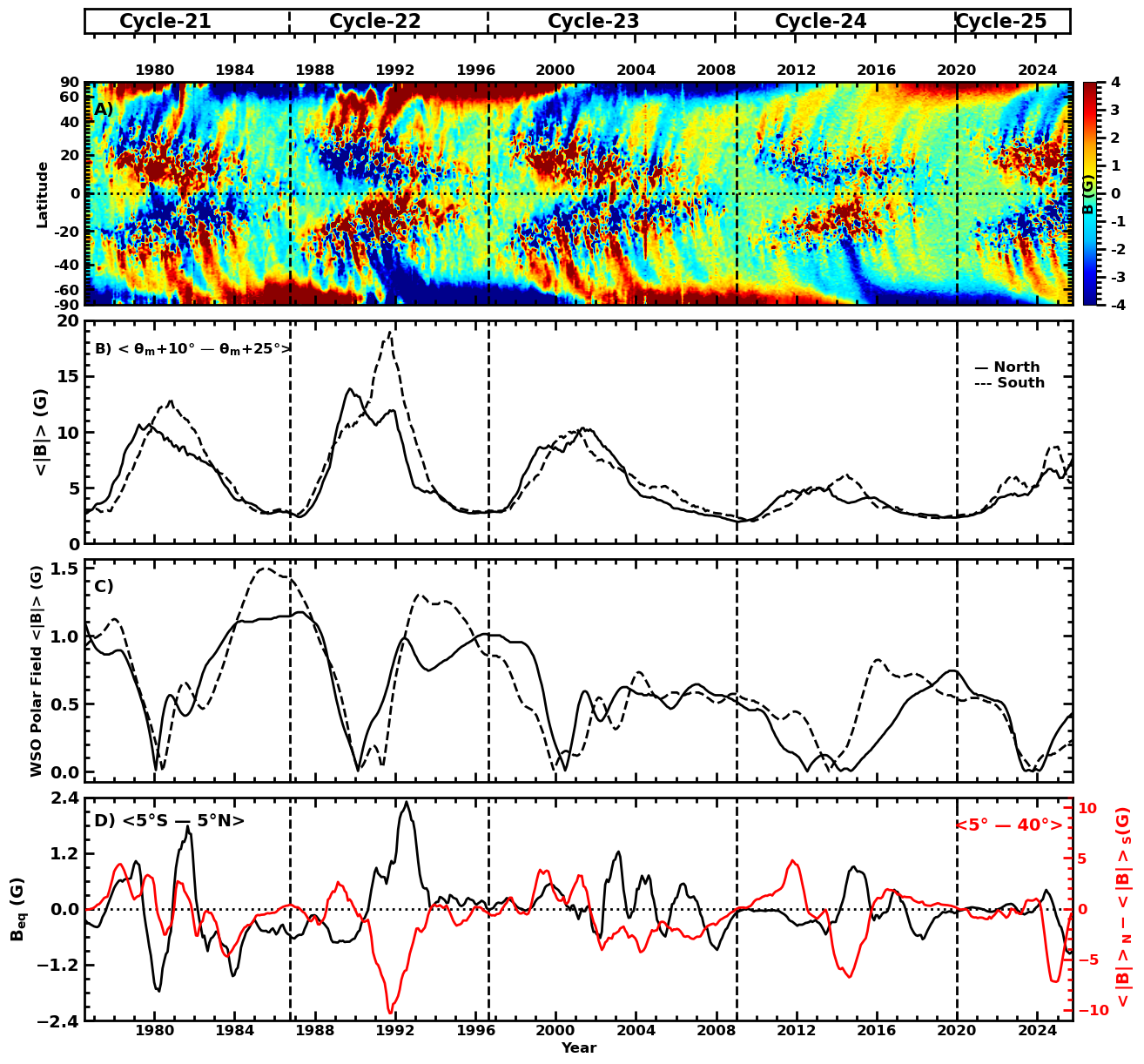}{1\textwidth}{}
          }
\caption{Panel A shows the temporal and latitudinal evolution of longitudinally averaged signed magnetic field 
from solar cycle 21 to the early phase of cycle 25, based on NSO Kitt Peak, SOLIS, and HMI synoptic magnetograms. Panel B 
shows the unsigned magnetic field ($|B|$) averaged over 10°-25° away from the mean latitude, $\theta_{m}$, of peak magnetic 
flux in each hemisphere, while panel C shows the unsigned polar field from WSO measurements. In panel D, the black solid curve is the north-south average of signed magnetic field intensity within 5°S-5°N (left y-axis), B$_{eq}$, that represents
the cross-equatorial flux plumes, while the red curve is the hemispheric asymmetry of active region magnetic field, B$_{asym}$,
calculated as the difference between the absolute magnetic field averaged over active latitudes ($5 - 40^{\circ}$)
in the north and south.} 
\label{fig:3}
\end{figure*}

Comparing meridional flow residuals and the magnetic butterfly diagram in Figure \ref{fig:1}, contrary 
to widely modelled effects of near-surface inflows \citep{2010ApJ...717..597J,2012A&A...548A..57C,2014ApJ...780....5U},
we do not observe any correlation between such flows (top panel of Figure \ref{fig:1}) and reduction in poleward 
transport of magnetic flux. Rather, we observe that the episodic poleward transport of magnetic fields is closely 
linked with the outflows at the base of the NSSL (depth of 0.95 $R_{\sun}$, Panel B): for example, during 2012–2013, the stonger 
poleward outflow (at 0.95 $R_{\sun}$) from active latitudes of the northern hemisphere (Panel B) coincides with enhanced 
poleward transport of magnetic flux (Panel C); likewise, during 2014–2016, the strongest poleward flux transport and outflow 
(at 0.95 $R_{\sun}$) episodes of Cycle 24 coincide in the southern hemisphere.

To examine the above connections more closely, we average the flow residuals $\delta U_{\theta}$ over latitude from 10° to 25° away from the
mean latitude, $\theta_{m}$, of active regions (determined as the latitude of peak flux at each time instant, in each hemisphere)
over the northern and southern hemispheres separately. The overplotted solid line in panel D of Figure \ref{fig:1} marks 
$\theta_{m}$, while the dot-dashed lines in panels A, B and D enclose the region 10° to 25° away from $\theta_{m}$ in both
hemispheres. The results for flows at the two depths are shown in panels A and B of Figure \ref{fig:2} [left panel], with solid and 
dashed lines for northern and southern hemispheres, respectively. Similarly averaged absolute values of magnetic field, 
representing the net flux transported to the poles, are plotted in panel C. And, we have also used the line-of-sight polar 
cap magnetic field measurements above 55° latitude from the Wilcox Solar Observatory (WSO), and plotted their unsigned amplitudes
in panel D of Figure \ref{fig:2}. We find that during 2014, the flow speed in the southern hemisphere at a depth of 
0.95 $R_{\sun}$ (red dashed curve in panel B of Figure \ref{fig:2}) reached its maximum value of about 12 m/s. This pronounced 
outflow exhibits a clear association with the magnetic flux over the same latitude range and time plotted as the 
dashed magenta curve in panel C of Figure \ref{fig:2}. The resulting poleward transport cause the subsequent accumulation 
of magnetic flux in the southern pole around 2016, approximately two years later, as shown by the black dashed curve 
in panel D of Figure \ref{fig:2}. In contrast, the corresponding poleward accumulation in the northern hemisphere occurred 
much later around 2020. The background meridional flow, which is predominantly poleward throughout the NSSL during cycle minima, 
facilitates the transport of magnetic flux from mid-latitudes toward the poles. During cycle maxima, the outflows beneath
active regions, at depths below 0.97 $R_{\sun}$, drive the early flux transport toward the poles. Therefore, subsurface 
outflows are a major driver of the flux transport process: higher flow velocities result in earlier flux arrival at the poles. 
Our results thus show that the near-surface inflows towards active regions do not significantly alter the flux transport, 
instead they promote the earlier buildup of polar flux via the deeper outflows to which they connect to.

During cycle 24 minimum (around the year 2020), a clear polar magnetic field asymmetry is observed (Figure \ref{fig:2}, panel D):
the accumulated flux is significantly higher in the northern hemisphere. The early build-up that led to the southern polar
field peaking in the year 2016 however declined in the subsequent years, likely due to the transport of opposite polarity
trailing field of bipolar regions with anti-Hale tilt \citep{2019SoPh..294...21M} (or the Hale's leading polarity of the hemisphere). 
In Figure \ref{fig:1} (panel C), several surges of leading polarity flux transported toward the pole in the southern hemisphere, 
during 2016 - 2019, are seen as due to a relative increase of spot regions that violated the Joy's law (see Figure 1 of 
\citet{2019SoPh..294...21M} that identifies anti-Joy and anti-Hale spots from Cycle 21 to 24). These events contributed to 
the overall weakening of the polar field in the southern hemisphere after 2016. Similarly, a year before Cycle 24 maximum 
(during 2013), just when the north pole was seeing a reversal with positive polarity flux replacing previous cycle's negative 
flux, the active latitudes in the north hosted a few instances of leading polarity negative flux transported towards the pole. 
This brief burst of negative flux nearly cancelled the already reversed pole 
around 2014, following which a subsequent second reversal happened when normal following polarity flux transport resumed in the north. 
Therefore, the subsurface outflows beneath active regions, which have surface inflows, drive the flux-transport process 
regardless of whether the spots are Hale or anti-Hale. Hale spots contribute to building up the polar field, while anti-Hale 
spots contribute to cancelling and thereby weakening the polar flux.

Implicit in our above inferences -- that flows at depths below 0.97 $R_{\sun}$ advectively transport magnetic flux --
is the assumption of high plasma-$\beta$ conditions, wherein the kinetic energy of flows exceeds the magnetic energy, by roughly 
an order of magnitude, at the spatial scales considered. We return to this point in detail in Section \ref{sec: conclusion}. 
We emphasize, however, that our measurements do not directly establish a causal relationship between the flows and flux transport. 
The inferred flows are from longitudinally averaged travel times and hence cannot be used to track their evolution in 
and around individual active regions. For the same reasons, our analysis does not allow us to probe the origin or driving 
mechanisms of these flows. 

Despite the above physical considerations, to examine the statistical significance of correlations between flows and
magnetic fields, we plot the scatter between them for the active phase of Cycle 24 (Oct 2010 - June 2018), with estimates of 
Spearman rank correlation coefficients, in the right-side panels of Figure \ref{fig:2}. The top four panels correspond to both 
quantities averaged over the same latitudinal range (10°-25° away from $\theta_{m}$), at depths of 0.95$R_{\odot}$ 
(left column) and 0.99$R_{\odot}$ (right column), for the northern (top row) and southern (bottom row) hemispheres, while
the bottom four panels show the correlation of the same flow signals against the unsigned magnetic field now averaged over
the whole active latitude range of 5°-40°. A stronger positive correlation is observed between the magnetic field and the 
outflow at the deeper layer (0.95 $R_{\sun}$), whereas at 0.99 $R_{\sun}$, since the inflows are directed opposite to the flux
transport toward higher latitutes, a strong negative correlation between them. Importantly, we find that the correlation 
coefficient increases when the magnetic field is averaged over the full active latitude range (5–40°; see the bottom four 
panels on the right of Figure \ref{fig:2}). This signifies that the whole flow system, of course, is due to the active region 
magnetic fields themselves, likely due to their interference with the flow of convective energy or thermal causes although 
our measurements and correlations studied here cannot address them. The ensuing flows thus draw their energy from the disturbed
thermal balance, leading to the dominance of kinetic energy of flows, which then take over the break-up and transport of
magnetic flux. We discuss this further in Section \ref{sec: conclusion}.

\subsection{Subsurface Flow Signatures in the Past Four Solar Cycles}
\label{subsec: results_2}
Having identified the roles of active-region driven meridional flow variations in the magnetic flux transport for
Cycle 24, we here examine the magnetic butterfly diagrams of the past three Cycles -- from Cycle 21 to 23 --
together with Cycle 24 and rising phase of Cycle 25. Panel A in Figure \ref{fig:3} shows the temporal and latitudinal
evolution of the signed magnetic field from Cycle 21 to early phase of Cycle 25, derived from NSO Kitt
Peak, SOLIS, and HMI synoptic magnetograms. Panel B shows the unsigned magnetic fields averaged over latitudes from 10° 
to 25° away from the mean latitude ($\theta_{m}$) of active regions in the northern (solid curve) and southern (dashed curve)
hemispheres. Drawing parralles with the connections between flows and magnetic flux 
transport shown in Figure \ref{fig:2}, we can identify instances of enhanced flux values with those of ouflows in the
lower half of the NSSL. Panel C shows the unsigned polar fields from the WSO. Comparing panels B and C, we can clearly identify 
the connections between the north-south asymmetry in sub-surface outflow driven flux transport and that of the polar field 
build-up, with a time-delay of two to three years (see Cycles 21, 22, and 24). Cycle 23 polar fields, despite significant
asymmetry in active region flux, had very little asymmetry and also were much weakened, which we identify as due to significant 
contributions from large non-Hale oriented spots in flux transport across the equator as well as towards the poles. 

\subsubsection{Cross-equator Flows, Flux Plumes, and Polar Fields}
An important consequence of flow circulations flanking the active regions is their cross-equatorial excursions
during cycle maxima when spots are relatively closer to the equator with significant hemispheric asymmetry. Such
flows transport flux across the equator leading to cross-equatorial flux plumes. A detailed analysis of such
cross-equator flows and flux plumes over the Cycle 24 was presented in our recent publication \citep{2026ApJ...997...57S}.
Here, we extend such analysis covering solar cycles 21 to the beginning of cycle 25.
Panel D of Figure \ref{fig:3} shows the north-south average of signed magnetic field
within 5°S-5°N (left y-axis, black), B$_{eq}$, representing the cross-equatorial flux plumes, while the right
y-axis has the corresponding hemispheric asymmetry of active region magnetic field estimated by taking the difference 
between the absolute magnetic field averaged over active latitudes ($5 - 40^{\circ}$)
in the north and south, B$_{asym}$=$< \vert$B$\vert_{N}>$ - $<\vert$B$\vert_{S}>$. 
The magnetic N-S asymmetry at active latitudes (red curve) provides a reliable
indication of the direction of cross-equatorial flows within the near-surface shear layer (NSSL).
For Cycle 24, we observed that, during peak activity, most of the flux plumes originated in the magnetically 
dominant southern hemisphere and migrated toward the northern hemisphere. Interestingly, these plumes moved 
counter to the direction of near-surface cross-equator flows. This seemingly paradoxical behaviour was explained 
by \citet{2026ApJ...997...57S} as due to the return flow (or outflows) at deeper layers (below 0.97 $R_{\sun}$), 
which actually carry magnetic flux across the equator in the reverse direction. 
The cross-equatorial flux plumes observed across all cycles (bottom panel of Figure \ref{fig:3})
show the same connection to north–south magnetic asymmetry at active latitudes (red curve in bottom panel). 
The flux plumes generally migrate from the more active hemisphere toward the less active
one. Because of the sign convention in our definition of the hemispheric magnetic asymmetry (B$_{asym}$), 
the above identified correlation between B$_{eq}$ and B$_{asym}$ should alternate between cycles as magnetic polarities
reverse from cycle to cycle (Hale's polarity rule): odd cycles would show positive correlation while the even ones have 
negative correlation. Panel D of Figure \ref{fig:3} largely confirms this connection between B$_{eq}$ and B$_{asym}$, however,
with a significant deviation in Cycle 23 (during the years mid-2002 to 2008). We identify this as due to
sunspots with non-Hale orientation appearing close to the equator and with significant hemispheric asymmetry. 
We in fact identify a large flux plume of positive polarity that crossed the equator from south to north, in September 2002 (see 
Panel A in Figure \ref{fig:3}), while the leading polarity in the south as per Hale's polarity rule is negative. 
Since the southern hemisphere was more active during this period, we expect a southward cross-equator flow near the surface
with a deeper northward return flow, which actually transports the positive flux of above identified non-Hale oriented spot.
 
In accordance with the expectation that the flux cancelled at the equator relates to the global dipole 
field \citep{2013A&A...557A.141C}, we find a strong relationship between the cross-equatorial flux plumes and the evolution 
of polar field asymmetry (Panels C and D of Figure \ref{fig:3}). In many instances, the emergence of these plumes 
appears to modulate the imbalance between the polar fields of the two hemispheres. 
Typically, the polar field in the hemisphere from which the plume originates strengthens relative to that of the 
opposite hemisphere. Furthermore, the magnitude of magnetic flux transported 
across the equator generally correlates well, in magnitude and time, with the enhancement of the polar field in the source hemisphere, which
leads the opposite one in the buildup of polar flux. Although this behaviour 
persists throughout all solar cycles, it becomes particularly evident during the periods around 1984, 1993 and 2015 for cycles 21, 22 and 24: the 
cross-equatorial flux plumes are of negative, positive and positive polarity, respectively, and the source hemisphere in all these cases is the southern one with corresponding increases in the southern polar field with very short time lags.

\section{Discussion, Conclusions and Broader Implications} \label{sec: conclusion}
In this work, we have used 14 years of time-distance helioseismic measurements from SDO/HMI, together with 
long-term synoptic magnetic field observations spanning four solar cycles, to establish a direct dynamical 
connection between active-region-driven subsurface flows in the NSSL and the global 
transport of magnetic flux that governs polar field buildup. Our results demonstrate that the episodic poleward and 
cross-equatorial transport of magnetic flux are primarily regulated by outflows 
originating beneath active latitudes in the lower half of the NSSL, rather than by near-surface inflows alone
\citep{2010ApJ...717..597J,2012A&A...548A..57C,2025A&A...701A.277C}.

A central result of this study is that enhanced poleward surges of magnetic flux coincide systematically with 
strengthened poleward-directed outflows at depths around 0.95$R_{\sun}$. These deeper outflows precede the arrival of 
flux at the poles by one to several years and largely determine the timing and amplitude of polar field buildup. 
This behavior is clearly manifested during Solar Cycle 24, where stronger subsurface outflows in the southern 
hemisphere during 2013-2015 led to an earlier accumulation of polar flux and a premature southern polar field 
maximum around 2016, while the northern hemisphere reached its polar maximum much later, near the cycle minimum 
in 2020. We have also verified such connections between flows, hemispheric magnetic asymmetry and polar fields
using synoptic magnetograms and WSO measurements covering Cycles 21 to 25.

Parameterized SFT models showed that converging inflows toward active regions can reduce the effective dipole 
moment of emerging bipoles and thereby weaken polar fields \citep{2010ApJ...717..597J},
while self-consistent implementations of magnetic-field–dependent inflows demonstrated that such flows provide 
a nonlinear feedback capable of regulating cycle amplitudes and reproducing the observed correlation between 
polar fields at solar minimum and the strength of the subsequent cycle \citep{2012A&A...548A..57C}. 
At the same time, advection-dominated SFT simulations driven by observed flows established that meridional-flow 
variations alone cannot explain the weak polar fields preceding Cycle 24, highlighting the dominant role of 
active-region source properties when cycles are intrinsically weak \citep{2014ApJ...780....5U}.

Our helioseismic results provide for a critical depth-resolved extension of SFT framework. By demonstrating that 
activity-modulated meridional flows are organized as a vertically structured circulation within the 
NSSL - characterized by near-surface inflows overlying deeper outflows - we provide direct observational support for the 
physical basis of flow-driven feedback mechanisms inferred in SFT models. Although previous helioseismic studies identified 
that converging inflows toward active regions near the surface and deeper outflows form circulation cells 
\citep{2002ApJ...570..855H,2004SoPh..220..371H,hindman2009subsurface}, our global scale flow measurements 
establish the dynamical coupling of deeper outflows and large-scale flux transport. 
An important physical basis for such deep-layer advection emerging as a dominant mechanism of flux transport derives 
from the high plasma $\beta$ conditions at deeper layers, where the deep-rooted flux is more passively advected
than at the surface layers. Such a scenerio also emerges from consideration of energetics on the spatial scales that
our measurements correspond to: for the Standard Solar Model S \citep{1996Sci...272.1286C}, kinetic energy density, 
$\rho v^{2}/2$, of typical poleward flows of magnitude of 15 m/s flanking the active latitudes at depths of $0.95R_{\sun}$, 
is about 900 $J/m^{3}$, which requires field strengths of about 480 G to react back on the flows or to be in equipartition. 
Since the surface (photospheric) fields are of strength about 10 - 20 G (\textit{c.f.} panel D Figure \ref{fig:1}),  
intensification of this field by more than a factor of 20, over the depth of 35 Mm, is required for it to reach equipartition,
which is highly unrealistic. Three-dimensional MHD simulations of sunspots with 3 kG fields at the photosphere typically 
achieve stable solutions with about 10 - 15 kG at depths of 15 - 20 Mm \citep{2011ApJ...740...15R}, a factor of 4 - 5 
increase in strength. Thus, although the magnetic energy densities at the core regions of active region flux are in 
super-equipartition to influence the energy transport and thereby cause thermally induced flow systems, the ensuing flows 
that spread over larger scales hold much larger kinetic energy to advect the disintegrating magnetic flux passively.
 

Our results are also consistent with, and provide a unifying physical interpretation for, earlier helioseismic and 
surface-magnetic-field studies that highlighted apparent discrepancies between surface meridional-flow measurements and magnetic flux 
transport. Using magnetic butterfly diagrams, \citet{2007ApJ...670L..69S} showed that the effective poleward flux-transport speed
matched better with the mean speed of meridional flows over depths of 3.5 - 12 Mm inferred from helioseismology, and
concluded that the localized flow perturbations around active regions strongly influence longitudinally averaged 
surface flows, which do not proportionally affect the global transport of magnetic flux. In parallel, \citet{2014ApJ...789L...7Z} 
reported an anti-correlation between high-latitude meridional-flow speed and the polarity of poleward-moving magnetic flux
during the rising phase of Solar Cycle 24, suggesting a dynamic coupling between surface flows and polar-field evolution. 
The depth-resolved and long-term helioseismic measurements presented here reconcile these findings by demonstrating that 
such surface-level signatures are the manifestation of a vertically structured flow system within the near-surface shear layer.

Our analysis further demonstrates that hemispheric asymmetries in polar field buildup are dynamically regulated by 
asymmetric subsurface flows, including cross-equatorial components. Periods of strong north–south imbalance in 
active-region magnetic flux give rise to cross-equatorial outflows in the deeper NSSL, which transport magnetic flux 
preferentially from the more active hemisphere toward the less active one. Such hemispheric coupling has long been 
inferred from surface magnetic field evolution (e.g., \citet{2009ApJ...707.1372W};\citet{2010ApJ...717..597J}), but our results 
identify a specific subsurface flow mechanism responsible for this coupling and establish its characteristic time 
delay of approximately two to three years. 

An important corollary is that the influence of anti-Hale and anti-Joy active regions on polar field evolution is 
mediated by the same subsurface flow system. While deeper outflows transport magnetic flux toward the poles largely 
independent of polarity, the contribution of this flux to the polar field depends sensitively on the polarity 
orientation and tilt of the emerging regions (e.g., \citet{2010A&A...518A...7D}; \citet{2014ApJ...791....5J}). This explains why 
strong subsurface outflows do not necessarily lead to sustained polar field strengthening, as demonstrated by the 
weakening of the southern polar field after 2016 despite high outflow velocities.

A further important aspect of subsurface outflows identified in this study concerns the production of Joy's law tilt itself,
which has recently been attributed to Coriolis force acting on supergranular-scale convective outflows \citep{2025A&A...700A..28R}. 
The Coriolis force mediation of active region flows, including its deeper outflows, which operate on larger spatial and 
temporal scales, have been well recognised and studied (see \citet{2025ApJ...984L...1S} and references therein). Hence,
although the deeper outflows should contribute to tilt production or modulate the subsequent redistribution of already tilted flux, 
whether they constitute a subsurface mechanism to generate tilt during the earliest stages of emergence remains an open 
question requiring high-cadence local helioseismic measurements that image the whole NSSL in latitude and longitude. 
This aspect relates to the origin of the whole circulatory flow system around active regions: whether they are driven 
from the surface layers after emergence, or due to processes related to the emerging flux itself as it traverses the 
deeper layers of the NSSL.

Taken together, these results point to a shallow but dynamically rich nonlinearity in the Babcock–Leighton dynamo, 
operating predominantly within the NSSL. The NSSL, thus, emerges as the critical interface where surface 
magnetic activity feeds back onto large-scale flows, which in turn regulate flux transport, hemispheric coupling, and 
polar field formation. This perspective reconciles previously divergent interpretations regarding the relative 
importance of flow variations and active-region source properties by identifying their complementary roles: 
stochastic variations in flux emergence govern cycle-to-cycle variability, while depth-dependent flow perturbations 
in the NSSL provide systematic nonlinear regulation \citep{2012A&A...548A..57C,2014ApJ...780....5U}.

The broader implication is that solar-cycle predictability is fundamentally controlled by processes occurring in 
the Sun’s outer few percent by radius. The success of polar-field–based predictions, including the correct anticipation of the 
stronger-than-expected Solar Cycle 25 (e.g., \citet{2016JGRA..12110744H}; \citet{2016ApJ...823L..22C}), reflects 
the limited memory of the solar dynamo and the dominant role of near-surface and subsurface transport processes. 
Future data-assimilative dynamo and surface flux transport models must therefore incorporate observationally constrained, 
depth-dependent meridional flows, including their hemispheric and activity-dependent variations, in order to capture 
the true dynamics of the solar magnetic cycle.

\section{Acknowledgements} 

The HMI data used are courtesy of NASA/SDO and the HMI science teams. Data preparation and processing have 
utilised the Data Record Management System (DRMS) software at the Joint Science Operations Center (JSOC) for 
NASA/SDO at Stanford University. Our sincere thanks go to H.M. Antia for supporting A.S. in performing the 
inversions. The GONG data used are obtained by the NSO Integrated Synoptic Program, managed by the National Solar 
Observatory, which is operated by the Association of Universities for Research in Astronomy (AURA), Inc., under a 
cooperative agreement with the National Science Foundation and with contributions from the National Oceanic and 
Atmospheric Administration. This work has received funding from the NASA DRIVE Science Center COFFIES Phase II 
CAN 80NSSC22M0162 to Stanford University. J.Z., R.C., and S.K. were partly sponsored by NASA HGIO grant 80NSSC25K7671. 
Data-intensive computations in this work have utilised the High-Performance Computing 
facility at the Indian   Institute of Astrophysics. A.S. is supported by INSPIRE Fellowship from the Department of 
Science and Technology (DST), Government of India. We thank an anonymous referee for valuable suggestions that led to significant improvements and additions to the manuscript.

\bibliography{msbib}{}

@ARTICLE{1996Sci...272.1286C,
       author = {{Christensen-Dalsgaard}, J. and {Dappen}, W. and {Ajukov}, S.~V. and {Anderson}, E.~R. and {Antia}, H.~M. and {Basu}, S. and {Baturin}, V.~A. and {Berthomieu}, G. and {Chaboyer}, B. and {Chitre}, S.~M. and {Cox}, A.~N. and {Demarque}, P. and {Donatowicz}, J. and {Dziembowski}, W.~A. and {Gabriel}, M. and {Gough}, D.~O. and {Guenther}, D.~B. and {Guzik}, J.~A. and {Harvey}, J.~W. and {Hill}, F. and {Houdek}, G. and {Iglesias}, C.~A. and {Kosovichev}, A.~G. and {Leibacher}, J.~W. and {Morel}, P. and {Proffitt}, C.~R. and {Provost}, J. and {Reiter}, J. and {Rhodes}, Jr., E.~J. and {Rogers}, F.~J. and {Roxburgh}, I.~W. and {Thompson}, M.~J. and {Ulrich}, R.~K.},
        title = "{The Current State of Solar Modeling}",
      journal = {Science},
         year = 1996,
        month = may,
       volume = {272},
       number = {5266},
        pages = {1286-1292},
          doi = {10.1126/science.272.5266.1286},
       adsurl = {https://ui.adsabs.harvard.edu/abs/1996Sci...272.1286C}
}

@ARTICLE{2015ApJ...808L..28J,
       author = {{Jiang}, J. and {Cameron}, R.~H. and {Sch{\"u}ssler}, M.},
        title = "{The Cause of the Weak Solar Cycle 24}",
      journal = {\apjl},
         year = 2015,
        month = jul,
       volume = {808},
       number = {1},
          eid = {L28},
        pages = {L28},
          doi = {10.1088/2041-8205/808/1/L28},
archivePrefix = {arXiv},
       eprint = {1507.01764},
 primaryClass = {astro-ph.SR},
       adsurl = {https://ui.adsabs.harvard.edu/abs/2015ApJ...808L..28J}
}

@ARTICLE{2017ApJ...849..144C,
       author = {{Chen}, Ruizhu and {Zhao}, Junwei},
        title = "{A Comprehensive Method to Measure Solar Meridional Circulation and the Center-to-limb Effect Using Time-Distance Helioseismology}",
      journal = {\apj},
         year = 2017,
        month = nov,
       volume = {849},
       number = {2},
          eid = {144},
        pages = {144},
          doi = {10.3847/1538-4357/aa8eec},
archivePrefix = {arXiv},
       eprint = {1709.07905},
 primaryClass = {astro-ph.SR},
       adsurl = {https://ui.adsabs.harvard.edu/abs/2017ApJ...849..144C}
}

@ARTICLE{2025ApJ...984L...1S,
       author = {{Sen}, Anisha and {Rajaguru}, S.~P. and {Iyer}, Abhinav Govindan and {Chen}, Ruizhu and {Zhao}, Junwei and {Kholikov}, Shukur},
        title = "{Solar Cycle Variations in Meridional Flows and Rotational Shear within the Sun's Near-surface Shear Layer}",
      journal = {\apjl},
         year = 2025,
        month = may,
       volume = {984},
       number = {1},
          eid = {L1},
        pages = {L1},
          doi = {10.3847/2041-8213/adc919},
       adsurl = {https://ui.adsabs.harvard.edu/abs/2025ApJ...984L...1S}
}

@ARTICLE{2010Sci...327.1350H,
       author = {{Hathaway}, David H. and {Rightmire}, Lisa},
        title = "{Variations in the Sun{\textquoteright}s Meridional Flow over a Solar Cycle}",
      journal = {Science},
         year = 2010,
        month = mar,
       volume = {327},
       number = {5971},
        pages = {1350},
          doi = {10.1126/science.1181990},
       adsurl = {https://ui.adsabs.harvard.edu/abs/2010Sci...327.1350H}
}

@ARTICLE{2023SSRv..219...31Y,
       author = {{Yeates}, Anthony R. and {Cheung}, Mark C.~M. and {Jiang}, Jie and {Petrovay}, Kristof and {Wang}, Yi-Ming},
        title = "{Surface Flux Transport on the Sun}",
      journal = {\ssr},
         year = 2023,
        month = jun,
       volume = {219},
       number = {4},
          eid = {31},
        pages = {31},
          doi = {10.1007/s11214-023-00978-8},
archivePrefix = {arXiv},
       eprint = {2303.01209},
 primaryClass = {astro-ph.SR},
       adsurl = {https://ui.adsabs.harvard.edu/abs/2023SSRv..219...31Y}
}

@ARTICLE{2013A&A...557A.141C,
       author = {{Cameron}, R.~H. and {Dasi-Espuig}, M. and {Jiang}, J. and {I{\c{s}}{\i}k}, E. and {Schmitt}, D. and {Sch{\"u}ssler}, M.},
        title = "{Limits to solar cycle predictability: Cross-equatorial flux plumes}",
      journal = {\aap},
         year = 2013,
        month = sep,
       volume = {557},
          eid = {A141},
        pages = {A141},
          doi = {10.1051/0004-6361/201321981},
archivePrefix = {arXiv},
       eprint = {1308.2827},
 primaryClass = {astro-ph.SR},
       adsurl = {https://ui.adsabs.harvard.edu/abs/2013A&A...557A.141C}
}

@ARTICLE{mahajan2023removal,
       author = {{Mahajan}, Sushant S. and {Sun}, Xudong and {Zhao}, Junwei},
        title = "{Removal of Active Region Inflows Reveals a Weak Solar Cycle Scale Trend in the Near-surface Meridional Flow}",
      journal = {\apj},
         year = 2023,
        month = jun,
       volume = {950},
       number = {1},
          eid = {63},
        pages = {63},
          doi = {10.3847/1538-4357/acc839},
archivePrefix = {arXiv},
       eprint = {2304.02158},
 primaryClass = {astro-ph.SR},
       adsurl = {https://ui.adsabs.harvard.edu/abs/2023ApJ...950...63M}
}

@article{hindman2009subsurface,
       author = {{Hindman}, Bradley W. and {Haber}, Deborah A. and {Toomre}, Juri},
        title = "{Subsurface Circulations within Active Regions}",
      journal = {\apj},
         year = 2009,
        month = jun,
       volume = {698},
       number = {2},
        pages = {1749-1760},
          doi = {10.1088/0004-637X/698/2/1749},
archivePrefix = {arXiv},
       eprint = {0904.1575},
 primaryClass = {astro-ph.SR},
       adsurl = {https://ui.adsabs.harvard.edu/abs/2009ApJ...698.1749H}
}

@article{rajaguru2015meridional,
       author = {{Rajaguru}, S.~P. and {Antia}, H.~M.},
        title = "{Meridional Circulation in the Solar Convection Zone: Time-Distance Helioseismic Inferences from Four Years of HMI/SDO Observations}",
      journal = {\apj},
         year = 2015,
        month = nov,
       volume = {813},
       number = {2},
          eid = {114},
        pages = {114},
          doi = {10.1088/0004-637X/813/2/114},
archivePrefix = {arXiv},
       eprint = {1510.01843},
 primaryClass = {astro-ph.SR},
       adsurl = {https://ui.adsabs.harvard.edu/abs/2015ApJ...813..114R}
}

@article{duvall1993time,
       author = {{Duvall}, Jr., T.~L. and {Jefferies}, S.~M. and {Harvey}, J.~W. and {Pomerantz}, M.~A.},
        title = "{Time-distance helioseismology}",
      journal = {\nat},
         year = 1993,
        month = apr,
       volume = {362},
       number = {6419},
        pages = {430-432},
          doi = {10.1038/362430a0},
       adsurl = {https://ui.adsabs.harvard.edu/abs/1993Natur.362..430D}
}

@article{scherrer2012helioseismic,
       author = {{Scherrer}, P.~H. and {Schou}, J. and {Bush}, R.~I. and {Kosovichev}, A.~G. and {Bogart}, R.~S. and {Hoeksema}, J.~T. and {Liu}, Y. and {Duvall}, T.~L. and {Zhao}, J. and {Title}, A.~M. and {Schrijver}, C.~J. and {Tarbell}, T.~D. and {Tomczyk}, S.},
        title = "{The Helioseismic and Magnetic Imager (HMI) Investigation for the Solar Dynamics Observatory (SDO)}",
      journal = {\solphys},
         year = 2012,
        month = jan,
       volume = {275},
       number = {1-2},
        pages = {207-227},
          doi = {10.1007/s11207-011-9834-2},
       adsurl = {https://ui.adsabs.harvard.edu/abs/2012SoPh..275..207S}
}

@article{antia2022changes,
 author = {{Antia}, H.~M. and {Basu}, Sarbani},
        title = "{Changes in the Near-surface Shear Layer of the Sun}",
      journal = {\apj},
         year = 2022,
        month = jan,
       volume = {924},
       number = {1},
          eid = {19},
        pages = {19},
          doi = {10.3847/1538-4357/ac32c3},
archivePrefix = {arXiv},
       eprint = {2110.13952},
 primaryClass = {astro-ph.SR},
       adsurl = {https://ui.adsabs.harvard.edu/abs/2022ApJ...924...19A}
}

@ARTICLE{2004SoPh..220..371H,
       author = {{Haber}, D.~A. and {Hindman}, B.~W. and {Toomre}, J. and {Thompson}, M.~J.},
        title = "{Organized Subsurface Flows near Active Regions}",
      journal = {\solphys},
         year = 2004,
        month = apr,
       volume = {220},
       number = {2},
        pages = {371-380},
          doi = {10.1023/B:SOLA.0000031405.52911.08},
       adsurl = {https://ui.adsabs.harvard.edu/abs/2004SoPh..220..371H}
}

@ARTICLE{2010ApJ...720.1030C,
       author = {{Cameron}, R.~H. and {Sch{\"u}ssler}, M.},
        title = "{Changes of the Solar Meridional Velocity Profile During Cycle 23 Explained by Flows Toward the Activity Belts}",
      journal = {\apj},
         year = 2010,
        month = sep,
       volume = {720},
       number = {2},
        pages = {1030-1032},
          doi = {10.1088/0004-637X/720/2/1030},
archivePrefix = {arXiv},
       eprint = {1007.2548},
 primaryClass = {astro-ph.SR},
       adsurl = {https://ui.adsabs.harvard.edu/abs/2010ApJ...720.1030C}
}

@ARTICLE{2024SoPh..299...42T,
       author = {{Teweldebirhan}, Kinfe and {Miesch}, Mark and {Gibson}, Sarah},
        title = "{Inflows Towards Bipolar Magnetic Active Regions and Their Nonlinear Impact on a Three-Dimensional Babcock{\textendash}Leighton Solar Dynamo Model}",
      journal = {\solphys},
         year = 2024,
        month = apr,
       volume = {299},
       number = {4},
          eid = {42},
        pages = {42},
          doi = {10.1007/s11207-024-02288-w},
archivePrefix = {arXiv},
       eprint = {2310.00738},
 primaryClass = {astro-ph.SR},
       adsurl = {https://ui.adsabs.harvard.edu/abs/2024SoPh..299...42T}
}

@ARTICLE{2019SoPh..294...21M,
       author = {{Mordvinov}, A.~V. and {Kitchatinov}, L.~L.},
        title = "{Evolution of the Sun's Polar Fields and the Poleward Transport of Remnant Magnetic Flux}",
      journal = {\solphys},
         year = 2019,
        month = feb,
       volume = {294},
       number = {2},
          eid = {21},
        pages = {21},
          doi = {10.1007/s11207-019-1410-1},
archivePrefix = {arXiv},
       eprint = {1902.00199},
 primaryClass = {astro-ph.SR},
       adsurl = {https://ui.adsabs.harvard.edu/abs/2019SoPh..294...21M}
}

@ARTICLE{2020ApJ...904...62W,
       author = {{Wang}, Zi-Fan and {Jiang}, Jie and {Zhang}, Jie and {Wang}, Jing-Xiu},
        title = "{Activity Complexes and a Prominent Poleward Surge during Solar Cycle 24}",
      journal = {\apj},
         year = 2020,
        month = nov,
       volume = {904},
       number = {1},
          eid = {62},
        pages = {62},
          doi = {10.3847/1538-4357/abbc1e},
archivePrefix = {arXiv},
       eprint = {2009.12483},
 primaryClass = {astro-ph.SR},
       adsurl = {https://ui.adsabs.harvard.edu/abs/2020ApJ...904...62W}
}

@ARTICLE{1996Sci...272.1300T,
       author = {{Thompson}, M.~J. and {Toomre}, J. and {Anderson}, E.~R. and {Antia}, H.~M. and {Berthomieu}, G. and {Burtonclay}, D. and {Chitre}, S.~M. and {Christensen-Dalsgaard}, J. and {Corbard}, T. and {De Rosa}, M. and {Genovese}, C.~R. and {Gough}, D.~O. and {Haber}, D.~A. and {Harvey}, J.~W. and {Hill}, F. and {Howe}, R. and {Korzennik}, S.~G. and {Kosovichev}, A.~G. and {Leibacher}, J.~W. and {Pijpers}, F.~P. and {Provost}, J. and {Rhodes}, Jr., E.~J. and {Schou}, J. and {Sekii}, T. and {Stark}, P.~B. and {Wilson}, P.~R.},
        title = "{Differential Rotation and Dynamics of the Solar Interior}",
      journal = {Science},
         year = 1996,
        month = may,
       volume = {272},
       number = {5266},
        pages = {1300-1305},
          doi = {10.1126/science.272.5266.1300},
       adsurl = {https://ui.adsabs.harvard.edu/abs/1996Sci...272.1300T}
}

@ARTICLE{2012A&A...548A..57C,
       author = {{Cameron}, R.~H. and {Sch{\"u}ssler}, M.},
        title = "{Are the strengths of solar cycles determined by converging flows towards the activity belts?}",
      journal = {\aap},
         year = 2012,
        month = dec,
       volume = {548},
          eid = {A57},
        pages = {A57},
          doi = {10.1051/0004-6361/201219914},
archivePrefix = {arXiv},
       eprint = {1210.7644},
 primaryClass = {astro-ph.SR},
       adsurl = {https://ui.adsabs.harvard.edu/abs/2012A&A...548A..57C}
}

@ARTICLE{2011Natur.471...80N,
       author = {{Nandy}, Dibyendu and {Mu{\~n}oz-Jaramillo}, Andr{\'e}s and {Martens}, Petrus C.~H.},
        title = "{The unusual minimum of sunspot cycle 23 caused by meridional plasma flow variations}",
      journal = {\nat},
         year = 2011,
        month = mar,
       volume = {471},
       number = {7336},
        pages = {80-82},
          doi = {10.1038/nature09786},
archivePrefix = {arXiv},
       eprint = {1303.0349},
 primaryClass = {astro-ph.SR},
       adsurl = {https://ui.adsabs.harvard.edu/abs/2011Natur.471...80N}
}

@ARTICLE{2008ApJ...688.1374T,
       author = {{Tsuneta}, S. and {Ichimoto}, K. and {Katsukawa}, Y. and {Lites}, B.~W. and {Matsuzaki}, K. and {Nagata}, S. and {Orozco Su{\'a}rez}, D. and {Shimizu}, T. and {Shimojo}, M. and {Shine}, R.~A. and {Suematsu}, Y. and {Suzuki}, T.~K. and {Tarbell}, T.~D. and {Title}, A.~M.},
        title = "{The Magnetic Landscape of the Sun's Polar Region}",
      journal = {\apj},
         year = 2008,
        month = dec,
       volume = {688},
       number = {2},
        pages = {1374-1381},
          doi = {10.1086/592226},
archivePrefix = {arXiv},
       eprint = {0807.4631},
 primaryClass = {astro-ph},
       adsurl = {https://ui.adsabs.harvard.edu/abs/2008ApJ...688.1374T}
}

@ARTICLE{2005GeoRL..32.1104S,
       author = {{Svalgaard}, Leif and {Cliver}, Edward W. and {Kamide}, Yohsuke},
        title = "{Sunspot cycle 24: Smallest cycle in 100 years?}",
      journal = {\grl},
         year = 2005,
        month = jan,
       volume = {32},
       number = {1},
          eid = {L01104},
        pages = {L01104},
          doi = {10.1029/2004GL021664},
       adsurl = {https://ui.adsabs.harvard.edu/abs/2005GeoRL..32.1104S}
}

@ARTICLE{2013ApJ...763...23S,
       author = {{Svalgaard}, Leif and {Kamide}, Yohsuke},
        title = "{Asymmetric Solar Polar Field Reversals}",
      journal = {\apj},
         year = 2013,
        month = jan,
       volume = {763},
       number = {1},
          eid = {23},
        pages = {23},
          doi = {10.1088/0004-637X/763/1/23},
archivePrefix = {arXiv},
       eprint = {1207.2077},
 primaryClass = {astro-ph.SR},
       adsurl = {https://ui.adsabs.harvard.edu/abs/2013ApJ...763...23S}
}

@ARTICLE{1978SoPh...58..225S,
       author = {{Svalgaard}, L. and {Duvall}, Jr., T.~L. and {Scherrer}, P.~H.},
        title = "{The strength of the Sun's polar fields.}",
      journal = {\solphys},
         year = 1978,
        month = jul,
       volume = {58},
       number = {2},
        pages = {225-239},
          doi = {10.1007/BF00157268},
       adsurl = {https://ui.adsabs.harvard.edu/abs/1978SoPh...58..225S}
}

@ARTICLE{1955ApJ...121..349B,
       author = {{Babcock}, Horace W. and {Babcock}, Harold D.},
        title = "{The Sun's Magnetic Field, 1952-1954.}",
      journal = {\apj},
         year = 1955,
        month = mar,
       volume = {121},
        pages = {349},
          doi = {10.1086/145994},
       adsurl = {https://ui.adsabs.harvard.edu/abs/1955ApJ...121..349B}
}

@ARTICLE{1959ApJ...130..364B,
       author = {{Babcock}, Harold D.},
        title = "{The Sun's Polar Magnetic Field.}",
      journal = {\apj},
         year = 1959,
        month = sep,
       volume = {130},
        pages = {364},
          doi = {10.1086/146726},
       adsurl = {https://ui.adsabs.harvard.edu/abs/1959ApJ...130..364B}
}

@ARTICLE{2010SoPh..267..267J,
       author = {{Janardhan}, P. and {Bisoi}, Susanta K. and {Gosain}, S.},
        title = "{Solar Polar Fields During Cycles 21 - 23: Correlation with Meridional Flows}",
      journal = {\solphys},
         year = 2010,
        month = dec,
       volume = {267},
       number = {2},
        pages = {267-277},
          doi = {10.1007/s11207-010-9653-x},
archivePrefix = {arXiv},
       eprint = {1009.4299},
 primaryClass = {physics.space-ph},
       adsurl = {https://ui.adsabs.harvard.edu/abs/2010SoPh..267..267J}
}

@ARTICLE{2010ApJ...717..597J,
       author = {{Jiang}, J. and {I{\c{s}}ik}, E. and {Cameron}, R.~H. and {Schmitt}, D. and {Sch{\"u}ssler}, M.},
        title = "{The Effect of Activity-related Meridional Flow Modulation on the Strength of the Solar Polar Magnetic Field}",
      journal = {\apj},
         year = 2010,
        month = jul,
       volume = {717},
       number = {1},
        pages = {597-602},
          doi = {10.1088/0004-637X/717/1/597},
archivePrefix = {arXiv},
       eprint = {1005.5317},
 primaryClass = {astro-ph.SR},
       adsurl = {https://ui.adsabs.harvard.edu/abs/2010ApJ...717..597J}
}

@ARTICLE{2011A&A...528A..83J,
       author = {{Jiang}, J. and {Cameron}, R.~H. and {Schmitt}, D. and {Sch{\"u}ssler}, M.},
        title = "{The solar magnetic field since 1700. II. Physical reconstruction of total, polar and open flux}",
      journal = {\aap},
         year = 2011,
        month = apr,
       volume = {528},
          eid = {A83},
        pages = {A83},
          doi = {10.1051/0004-6361/201016168},
archivePrefix = {arXiv},
       eprint = {1102.1270},
 primaryClass = {astro-ph.SR},
       adsurl = {https://ui.adsabs.harvard.edu/abs/2011A&A...528A..83J}
}

@ARTICLE{2014ApJ...780....5U,
       author = {{Upton}, Lisa and {Hathaway}, David H.},
        title = "{Predicting the Sun's Polar Magnetic Fields with a Surface Flux Transport Model}",
      journal = {\apj},
         year = 2014,
        month = jan,
       volume = {780},
       number = {1},
          eid = {5},
        pages = {5},
          doi = {10.1088/0004-637X/780/1/5},
archivePrefix = {arXiv},
       eprint = {1311.0844},
 primaryClass = {astro-ph.SR},
       adsurl = {https://ui.adsabs.harvard.edu/abs/2014ApJ...780....5U}
}

@ARTICLE{2024RAA....24g5015Y,
       author = {{Yang}, Shuhong and {Jiang}, Jie and {Wang}, Zifan and {Hou}, Yijun and {Jin}, Chunlan and {Song}, Qiao and {Luo}, Yukun and {Li}, Ting and {Zhang}, Jun and {Zhang}, Yuzong and {Zhou}, Guiping and {Deng}, Yuanyong and {Wang}, Jingxiu},
        title = "{Long-term Variation of the Solar Polar Magnetic Fields at Different Latitudes}",
      journal = {Research in Astronomy and Astrophysics},
         year = 2024,
        month = jul,
       volume = {24},
       number = {7},
          eid = {075015},
        pages = {075015},
          doi = {10.1088/1674-4527/ad539a},
archivePrefix = {arXiv},
       eprint = {2408.15168},
 primaryClass = {astro-ph.SR},
       adsurl = {https://ui.adsabs.harvard.edu/abs/2024RAA....24g5015Y}
}

@ARTICLE{2015ApJ...805..165L,
       author = {{Liang}, Zhi-Chao and {Chou}, Dean-Yi},
        title = "{Effects of Solar Surface Magnetic Fields on the Time-Distance Analysis of Solar Subsurface Meridional Flows}",
      journal = {\apj},
         year = 2015,
        month = jun,
       volume = {805},
       number = {2},
          eid = {165},
        pages = {165},
          doi = {10.1088/0004-637X/805/2/165},
       adsurl = {https://ui.adsabs.harvard.edu/abs/2015ApJ...805..165L}
}

@ARTICLE{1961ApJ...133..572B,
       author = {{Babcock}, H.~W.},
        title = "{The Topology of the Sun's Magnetic Field and the 22-Year Cycle.}",
      journal = {\apj},
         year = 1961,
        month = mar,
       volume = {133},
        pages = {572},
          doi = {10.1086/147060},
       adsurl = {https://ui.adsabs.harvard.edu/abs/1961ApJ...133..572B}
}

@ARTICLE{1964ApJ...140.1547L,
       author = {{Leighton}, Robert B.},
        title = "{Transport of Magnetic Fields on the Sun.}",
      journal = {\apj},
         year = 1964,
        month = nov,
       volume = {140},
        pages = {1547},
          doi = {10.1086/148058},
       adsurl = {https://ui.adsabs.harvard.edu/abs/1964ApJ...140.1547L}
}

@ARTICLE{1969ApJ...156....1L,
       author = {{Leighton}, Robert B.},
        title = "{A Magneto-Kinematic Model of the Solar Cycle}",
      journal = {\apj},
         year = 1969,
        month = apr,
       volume = {156},
        pages = {1},
          doi = {10.1086/149943},
       adsurl = {https://ui.adsabs.harvard.edu/abs/1969ApJ...156....1L}
}

@ARTICLE{2014ARA&A..52..251C,
       author = {{Charbonneau}, Paul},
        title = "{Solar Dynamo Theory}",
      journal = {\araa},
         year = 2014,
        month = aug,
       volume = {52},
        pages = {251-290},
          doi = {10.1146/annurev-astro-081913-040012},
       adsurl = {https://ui.adsabs.harvard.edu/abs/2014ARA&A..52..251C}
}

@ARTICLE{1995A&A...303L..29C,
       author = {{Choudhuri}, A.~R. and {Schussler}, M. and {Dikpati}, M.},
        title = "{The solar dynamo with meridional circulation.}",
      journal = {\aap},
         year = 1995,
        month = nov,
       volume = {303},
        pages = {L29},
       adsurl = {https://ui.adsabs.harvard.edu/abs/1995A&A...303L..29C}
}

@ARTICLE{1991ApJ...375..761W,
       author = {{Wang}, Y.-M. and {Sheeley}, Jr., N.~R.},
        title = "{Magnetic Flux Transport and the Sun's Dipole Moment: New Twists to the Babcock-Leighton Model}",
      journal = {\apj},
         year = 1991,
        month = jul,
       volume = {375},
        pages = {761},
          doi = {10.1086/170240},
       adsurl = {https://ui.adsabs.harvard.edu/abs/1991ApJ...375..761W}
}

@ARTICLE{2025A&A...701A.277C,
       author = {{Cameron}, R.~H. and {Schunker}, H. and {Brun}, A.~S. and {Strugarek}, A. and {Finley}, A.~J. and {Roland-Batty}, W. and {Birch}, A.~C. and {Gizon}, L.},
        title = "{Closing the solar dynamo loop: Poloidal field generated at the surface by plasma flows}",
      journal = {\aap},
         year = 2025,
        month = sep,
       volume = {701},
          eid = {A277},
        pages = {A277},
          doi = {10.1051/0004-6361/202553844},
       adsurl = {https://ui.adsabs.harvard.edu/abs/2025A&A...701A.277C}
}

@ARTICLE{2023SSRv..219...39H,
       author = {{Hazra}, Gopal and {Nandy}, Dibyendu and {Kitchatinov}, Leonid and {Choudhuri}, Arnab Rai},
        title = "{Mean Field Models of Flux Transport Dynamo and Meridional Circulation in the Sun and Stars}",
      journal = {\ssr},
         year = 2023,
        month = aug,
       volume = {219},
       number = {5},
          eid = {39},
        pages = {39},
          doi = {10.1007/s11214-023-00982-y},
archivePrefix = {arXiv},
       eprint = {2302.09390},
 primaryClass = {astro-ph.SR},
       adsurl = {https://ui.adsabs.harvard.edu/abs/2023SSRv..219...39H}
}

@ARTICLE{2014ApJ...785L...8M,
       author = {{Miesch}, Mark S. and {Dikpati}, Mausumi},
        title = "{A Three-dimensional Babcock-Leighton Solar Dynamo Model}",
      journal = {\apjl},
         year = 2014,
        month = apr,
       volume = {785},
       number = {1},
          eid = {L8},
        pages = {L8},
          doi = {10.1088/2041-8205/785/1/L8},
archivePrefix = {arXiv},
       eprint = {1401.6557},
 primaryClass = {astro-ph.SR},
       adsurl = {https://ui.adsabs.harvard.edu/abs/2014ApJ...785L...8M}
}

@ARTICLE{2025A&A...700A..28R,
       author = {{Roland-Batty}, W. and {Schunker}, H. and {Cameron}, R.~H. and {Przybylski}, D. and {Gizon}, L. and {Pontin}, D.~I.},
        title = "{Coriolis force acting on near-surface horizontal flows during simulations of flux emergence produces a tilt angle consistent with Joy's law on the Sun}",
      journal = {\aap},
         year = 2025,
        month = aug,
       volume = {700},
          eid = {A28},
        pages = {A28},
          doi = {10.1051/0004-6361/202554253},
archivePrefix = {arXiv},
       eprint = {2506.15935},
 primaryClass = {astro-ph.SR},
       adsurl = {https://ui.adsabs.harvard.edu/abs/2025A&A...700A..28R}
}

@ARTICLE{2007ApJ...670L..69S,
       author = {{{\v{S}}vanda}, Michal and {Kosovichev}, Alexander G. and {Zhao}, Junwei},
        title = "{Speed of Meridional Flows and Magnetic Flux Transport on the Sun}",
      journal = {\apjl},
         year = 2007,
        month = nov,
       volume = {670},
       number = {1},
        pages = {L69-L72},
          doi = {10.1086/524059},
archivePrefix = {arXiv},
       eprint = {0710.0590},
 primaryClass = {astro-ph},
       adsurl = {https://ui.adsabs.harvard.edu/abs/2007ApJ...670L..69S}
}

@ARTICLE{2009ApJ...707.1372W,
       author = {{Wang}, Y.-M. and {Robbrecht}, E. and {Sheeley}, Jr., N.~R.},
        title = "{On the Weakening of the Polar Magnetic Fields during Solar Cycle 23}",
      journal = {\apj},
         year = 2009,
        month = dec,
       volume = {707},
       number = {2},
        pages = {1372-1386},
          doi = {10.1088/0004-637X/707/2/1372},
       adsurl = {https://ui.adsabs.harvard.edu/abs/2009ApJ...707.1372W}
}

@ARTICLE{2014ApJ...789L...7Z,
       author = {{Zhao}, Junwei and {Kosovichev}, A.~G. and {Bogart}, R.~S.},
        title = "{Solar Meridional Flow in the Shallow Interior during the Rising Phase of Cycle 24}",
      journal = {\apjl},
         year = 2014,
        month = jul,
       volume = {789},
       number = {1},
          eid = {L7},
        pages = {L7},
          doi = {10.1088/2041-8205/789/1/L7},
archivePrefix = {arXiv},
       eprint = {1406.2735},
 primaryClass = {astro-ph.SR},
       adsurl = {https://ui.adsabs.harvard.edu/abs/2014ApJ...789L...7Z}
}

@ARTICLE{2026ApJ...997...57S,
       author = {{Sen}, Anisha and {Rajaguru}, S.~P. and {Chen}, Ruizhu and {Zhao}, Junwei and {Kholikov}, Shukur},
        title = "{Hemispheric Magnetic Asymmetry and Cross-equatorial Circulation Cells within the Sun's Near-surface Shear Layer}",
      journal = {\apj},
         year = 2026,
        month = jan,
       volume = {997},
       number = {1},
          eid = {57},
        pages = {57},
          doi = {10.3847/1538-4357/ae2c85},
archivePrefix = {arXiv},
       eprint = {2512.12327},
 primaryClass = {astro-ph.SR},
       adsurl = {https://ui.adsabs.harvard.edu/abs/2026ApJ...997...57S}
}

@ARTICLE{2002ApJ...570..855H,
       author = {{Haber}, Deborah A. and {Hindman}, Bradley W. and {Toomre}, Juri and {Bogart}, Richard S. and {Larsen}, Rasmus M. and {Hill}, Frank},
        title = "{Evolving Submerged Meridional Circulation Cells within the Upper Convection Zone Revealed by Ring-Diagram Analysis}",
      journal = {\apj},
         year = 2002,
        month = may,
       volume = {570},
       number = {2},
        pages = {855-864},
          doi = {10.1086/339631},
       adsurl = {https://ui.adsabs.harvard.edu/abs/2002ApJ...570..855H}
}

@ARTICLE{2010A&A...518A...7D,
       author = {{Dasi-Espuig}, M. and {Solanki}, S.~K. and {Krivova}, N.~A. and {Cameron}, R. and {Pe{\~n}uela}, T.},
        title = "{Sunspot group tilt angles and the strength of the solar cycle}",
      journal = {\aap},
         year = 2010,
        month = jul,
       volume = {518},
          eid = {A7},
        pages = {A7},
          doi = {10.1051/0004-6361/201014301},
archivePrefix = {arXiv},
       eprint = {1005.1774},
 primaryClass = {astro-ph.SR},
       adsurl = {https://ui.adsabs.harvard.edu/abs/2010A&A...518A...7D}
}

@ARTICLE{2014ApJ...791....5J,
       author = {{Jiang}, J. and {Cameron}, R.~H. and {Sch{\"u}ssler}, M.},
        title = "{Effects of the Scatter in Sunspot Group Tilt Angles on the Large-scale Magnetic Field at the Solar Surface}",
      journal = {\apj},
         year = 2014,
        month = aug,
       volume = {791},
       number = {1},
          eid = {5},
        pages = {5},
          doi = {10.1088/0004-637X/791/1/5},
archivePrefix = {arXiv},
       eprint = {1406.5564},
 primaryClass = {astro-ph.SR},
       adsurl = {https://ui.adsabs.harvard.edu/abs/2014ApJ...791....5J}
}

@ARTICLE{2016JGRA..12110744H,
       author = {{Hathaway}, David H. and {Upton}, Lisa A.},
        title = "{Predicting the amplitude and hemispheric asymmetry of solar cycle 25 with surface flux transport}",
      journal = {Journal of Geophysical Research (Space Physics)},
         year = 2016,
        month = nov,
       volume = {121},
       number = {11},
        pages = {10,744-10,753},
          doi = {10.1002/2016JA023190},
archivePrefix = {arXiv},
       eprint = {1611.05106},
 primaryClass = {astro-ph.SR},
       adsurl = {https://ui.adsabs.harvard.edu/abs/2016JGRA..12110744H}
}

@ARTICLE{2016ApJ...823L..22C,
       author = {{Cameron}, R.~H. and {Jiang}, J. and {Sch{\"u}ssler}, M.},
        title = "{Solar Cycle 25: Another Moderate Cycle?}",
      journal = {\apjl},
         year = 2016,
        month = jun,
       volume = {823},
       number = {2},
          eid = {L22},
        pages = {L22},
          doi = {10.3847/2041-8205/823/2/L22},
archivePrefix = {arXiv},
       eprint = {1604.05405},
 primaryClass = {astro-ph.SR},
       adsurl = {https://ui.adsabs.harvard.edu/abs/2016ApJ...823L..22C}
}

@ARTICLE{2011ApJ...740...15R,
       author = {{Rempel}, Matthias},
        title = "{Subsurface Magnetic Field and Flow Structure of Simulated Sunspots}",
      journal = {\apj},
         year = 2011,
        month = oct,
       volume = {740},
       number = {1},
          eid = {15},
        pages = {15},
          doi = {10.1088/0004-637X/740/1/15},
archivePrefix = {arXiv},
       eprint = {1106.6287},
 primaryClass = {astro-ph.SR},
       adsurl = {https://ui.adsabs.harvard.edu/abs/2011ApJ...740...15R}
}
\bibliographystyle{aasjournal}
\end{document}